\documentclass[sigconf,dvipsnames,screen,authorversion,nonacm]{acmart} 

\usepackage{multirow}
\usepackage{adjustbox}
\usepackage{listings}
\usepackage{subcaption}
\usepackage{todonotes}
\usepackage{float}

\usepackage{tikz}
\usetikzlibrary{fit,positioning}
\usetikzlibrary{calc}
\usepackage{multicol}

\newcommand{\inlineheadingbf}[1]{\medskip\noindent{\bfseries #1.}}

\usepackage[group-digits=integer, group-minimum-digits=4,
                list-final-separator={, and }, 
                free-standing-units, unit-optional-argument, 
                detect-weight=true, 
                ]{siunitx}
                
\newcommand{\realit}{RealiT}

\definecolor{correct-green}{HTML}{86AC41}

\lstdefinestyle{Python}{
	language=Python,
	basicstyle=\linespread{0.75}\small\ttfamily,
	columns=fullflexible,
	breaklines=true,
	tabsize=3,
	postbreak=\mbox{\textcolor{gray}{\tiny{$\rightarrow$}}\space},
	captionpos=b,
	numbers=left,
	linewidth=0.9\columnwidth,
	xleftmargin=2em,
	linebackgroundheight=0.8em,
	escapechar=$,
	commentstyle=\color{OliveGreen},
	keywordstyle=\color{RedViolet}\bfseries,
	stringstyle=\color{blue}\itshape,
	keywordstyle=[2]{\color{gray}},
	keywordstyle=[3]{\color{MidnightBlue}},
	morekeywords=[2]{@Override},
	morekeywords=[3]{first, second, list, sms, connection, privateDataManager, instances, a, b, result, entry, v, slot, start, _buf, CODE_W_SCOPE, name, temp, code, _code, _scope},
	belowcaptionskip=0cm,
}

\makeatletter
\let\old@lstKV@SwitchCases\lstKV@SwitchCases
\def\lstKV@SwitchCases#1#2#3{}
\makeatother
\usepackage{lstlinebgrd}
\makeatletter
\let\lstKV@SwitchCases\old@lstKV@SwitchCases

\lst@Key{numbers}{none}{%
	\def\lst@PlaceNumber{\lst@linebgrd}%
	\lstKV@SwitchCases{#1}%
	{none:\\%
		left:\def\lst@PlaceNumber{\llap{\normalfont
				\lst@numberstyle{\thelstnumber}\kern\lst@numbersep}\lst@linebgrd}\\%
		right:\def\lst@PlaceNumber{\rlap{\normalfont
				\kern\linewidth \kern\lst@numbersep
				\lst@numberstyle{\thelstnumber}}\lst@linebgrd}%
	}{\PackageError{Listings}{Numbers #1 unknown}\@ehc}}
\makeatother

\setcopyright{acmcopyright}
\copyrightyear{2018}
\acmYear{2018}
\acmDOI{XXXXXXX.XXXXXXX}

\acmConference[Preprint]{}{June 2022}{Woodstock, NY}

\begin{document}

\title{Can we learn from developer mistakes? \\ Learning to localize and repair
real bugs from real bug fixes}

\author{Cedric Richter}
\authornote{This author was partially supported by the German Research Foundation (DFG)
within the Collaborative Research Centre "On-The-Fly Computing" (SFB 901).}
\affiliation{
  \institution{University of Oldenburg}
  \streetaddress{Ammerländer Heerstraße 114-118}
  \city{Oldenburg}
  \country{Germany}
  \postcode{26129 }
}
\email{cedric.richter@uni-oldenburg.de}

\author{Heike Wehrheim}
\authornotemark[1]
\affiliation{
  \institution{University of Oldenburg}
  \streetaddress{Ammerländer Heerstraße 114-118}
  \city{Oldenburg}
  \country{Germany}
  \postcode{26129 }
}
\email{heike.wehrheim@uni-oldenburg.de}

\renewcommand{\shortauthors}{Richter and Wehrheim}

\begin{abstract}
Real bug fixes found in open source repositories seem to be the perfect
source for learning to localize and repair real bugs. However, the absence of large scale bug fix collections has made it difficult to effectively exploit real bug fixes in the training of larger neural models in the past.
In contrast, {\em artificial bugs} -- produced by mutating existing source code -- can be easily obtained at a sufficient scale and are therefore often preferred in the training of existing approaches. Still, localization and repair models that are trained on artificial bugs usually underperform when faced with real bugs. This raises the question whether bug localization and repair models trained on real bug fixes are more effective in localizing and repairing real bugs.

We address this question by introducing {\em \realit}, a pre-train-and-fine-tune approach for effectively learning to localize and repair real bugs from real bug fixes.
\realit~ is first pre-trained on a large number of artificial bugs produced by traditional mutation operators
and then fine-tuned on a smaller set of real bug fixes. Fine-tuning does not require any modifications of the learning
algorithm and hence can be easily adopted in various training scenarios for bug localization or repair (even when real training data is scarce). In addition, we found that training on real bug fixes with \realit~ is empirically powerful by nearly {\em doubling} the localization performance of an existing model on real bugs while maintaining or even improving the repair performance. 
\end{abstract}


\maketitle

\section{Introduction}
\begin{figure}[t]
		\begin{adjustbox}{max width=1\linewidth}
			\includegraphics{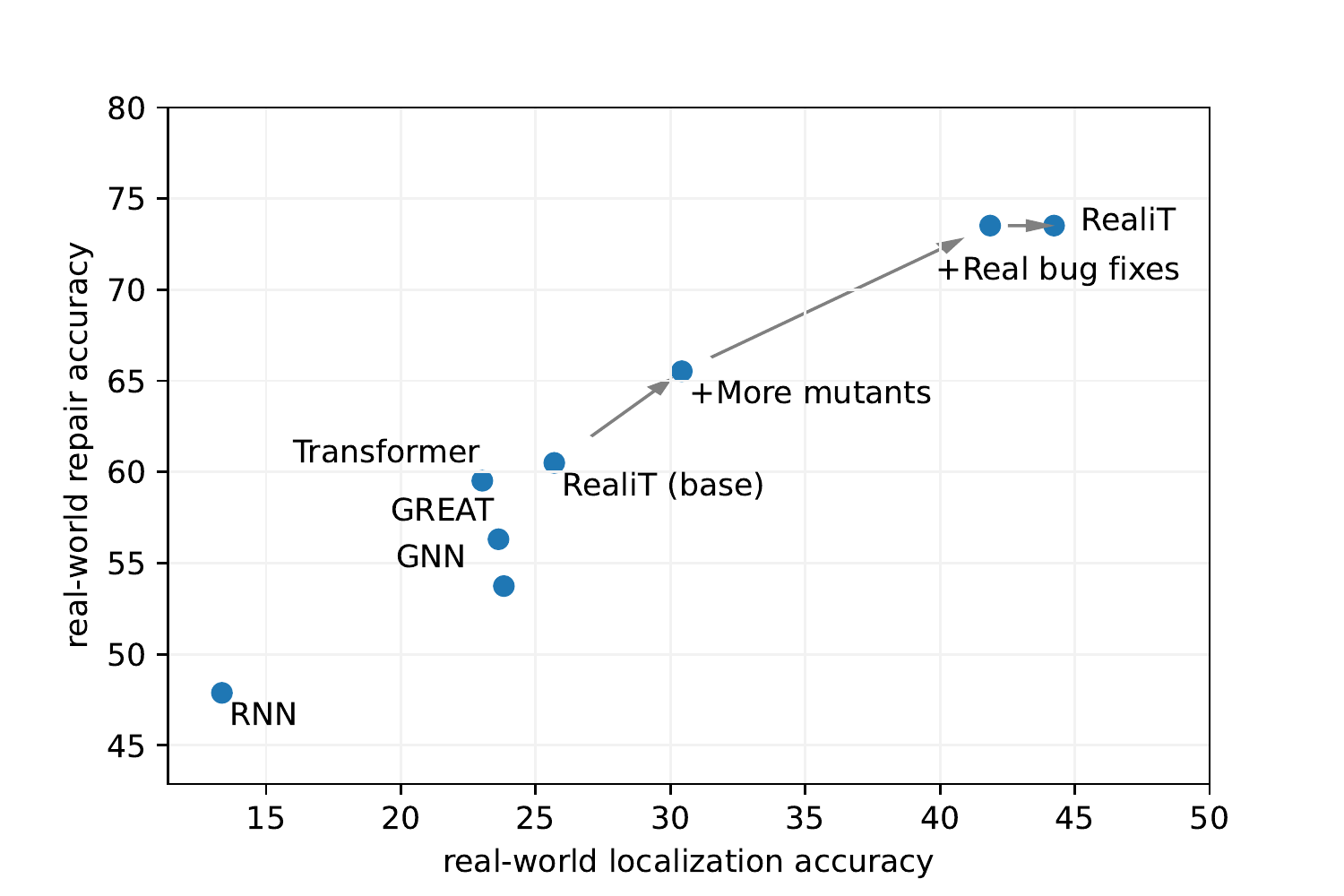}
		\end{adjustbox}
	\caption{Training on real bug fixes improves localization and repair of real bugs.}\label{fig:intro}
\end{figure}

Finding and fixing software bugs are one of the most common challenges in software engineering~\cite{DBLP:journals/corr/abs-2105-02162}. Developers are often faced with these tasks while spending a considerable amount of time in fixing software bugs. Still, some bugs find their way in open software repositories which then has to be fixed by a bug fixing code change. This raises the question whether we can relieve the developer from the debugging process by learning from common developer mistakes and their fixes found in open source projects.

Previous work~\cite{pradel2018deepbugs, hellendoorn2019global, vasic2019neural,allamanis2017learning, richter2022learning, allamanis2021self} addressed this question by designing automatic learning-based methods for bug localization and repair. However, to obtain the necessary amount of data needed for training, they often employed {\em code mutants} instead of real bug fixes. Mutants are generated by automatically injecting small code changes into existing code. As this process is automated, mutants can be easily obtained at large masses which is necessary for training effective learning-based models. However, a mutant may not represent a real bug which could ultimately bottleneck the performance of learning based bug localization repair. 

In contrast in this work, we aim to explore the effect of real bug fixes obtained from open source repositories on the training of learning-based bug localization and repair methods. 
For this, we employ a novel dataset of 33k real-world bug fixes obtained from public Python projects. 
Since this dataset is still comparably small to the datasets typically used for training localization and repair models~\cite{allamanis2021self, hellendoorn2019global} (which oftentimes contain millionth of artificial bugs), we propose \realit~(pronounced ``reality''), a novel training scheme for learning to \textbf{Re}pair \textbf{a}nd \textbf{l}ocal\textbf{i}ze with \textbf{T}ransformers. 
\realit~ is designed to combine the strengths of both training on mutants and real bug fixes by first pre-training on a high number of mutants and then fine-tuning on a smaller set of real bug fixes. This design allows us to not only evaluate the impact of real bug fixes and mutants together on the training process, but also individually (by skipping either pre-training or fine-tuning phase).

To evaluate the impact of \realit's training on the localization and repair of real bugs, we implement \realit~ for fixing a variety of {\em single token bugs} in Python. Our implementation
considers four common types of single token bugs -- that can be fixed by changing a single program token. We evaluate \realit~ together with several baselines on over 2000 real world bugs collected in the PyPIBugs benchmark~\cite{allamanis2021self}.

By integrating real bug fixes in the training process with \realit, we observe significant performance gains over a training that solely focuses on mutants (as in previous works). In fact, training with real bug fixes allows to nearly double the number of successfully localized real bugs (x-axis in Figure~\ref{fig:intro}) 
while also maintaining or improving the repair performance. 

Our main contributions can be summarized as follows:
\begin{itemize}
\item For investigating the effect of real bug fixes on the training of neural bug localization and repair models, we propose a simple pre-train-and-fine-tune approach. We find that training both on mutants and real bugs with our method significantly improves the performance over models solely trained on either mutants or real bugs when evaluated on real single token bugs in Python.
\item We show that data quality and quantity has a significant impact on neural bug localization and repair models. By pre-training on a large number of mutants (up to 20x larger than in previous work), \realit~ already significantly improves localization and repair performance both on mutants and real bugs. Combined with fine-tuning on real bug fixes, \realit~ is the first model to repair a significant portion of a real bugs benchmark.

\item For adopting \realit~ into future projects, we show that even training on smaller subsets of real bug fixes can yield performance improvement for localization and repair of real bugs. However, more bug fixes are also more beneficial.
\end{itemize}
We plan to release all trained models, pre-training and fine-tuning code\footnote{https://github.com/cedricrupb/nbfbaselines}.


\begin{table*}[ht!]
\caption{Examples of single token bug types taken from PyPIBugs~\cite{allamanis2021self}}
\label{tab:ex}
\begin{tabular}{ll}
\toprule
Example & Description\\
\midrule
 \begin{minipage}{\columnwidth}
\begin{lstlisting}[style=Python, breakindent=82.5px, linebackgroundcolor={\ifnum\value{lstnumber}=5 \color{red!15} \fi}]
# VarMisuse: $\textbf{\color{OliveGreen}applied}$ instead of $\textbf{\color{OliveGreen}patch}$
applied = self.db.applied_patches()
for patch in applied:
    if patch in patches:
        patches.remove($\textbf{\color{MidnightBlue}applied}$)

\end{lstlisting}
\end{minipage}
 & 
 \begin{minipage}{\columnwidth}
 All \texttt{applied} patches should be removed from the \texttt{patches} list. However, the developer mistakenly tries to remove \texttt{applied} instead of a single \texttt{patch}.\\
 \textbf{Fix:} replace \texttt{applied} in Line 5 by \texttt{patch} defined in Line 3.
 \end{minipage}
 \\
\midrule
 \begin{minipage}{\columnwidth}
\begin{lstlisting}[style=Python, breakindent=82.5px, linebackgroundcolor={\ifnum\value{lstnumber}=7 \color{red!15} \fi}]
# BinOp: $\textbf{\color{OliveGreen}!=}$ instead of $\textbf{\color{OliveGreen}==}$
def updateRefractionParameters(self):
    ...
    if self.ui.checkRefracNone.isChecked():
        return False
    if self.checkRefracNoTrack.isChecked():
        if self.app.mount.status $\textbf{\color{MidnightBlue}!=}$ 0:
            return False
     ...
\end{lstlisting}
\end{minipage}
 & 
 \begin{minipage}{\columnwidth}
The function \texttt{updateRefractionParameters} performs an update and returns true if the update was successful.
Prior to the update the function checks some preconditions and the function should abort if the \texttt{mount} is not ready.
Therefore,  we can conventionally expect that we abort if the status is zero. However, we check whether the status is not zero. \\
 \textbf{Fix:} replace \texttt{!=} in Line 7 by \texttt{==}.
 \end{minipage}
 \\
\midrule
 \begin{minipage}{\columnwidth}
\begin{lstlisting}[style=Python, breakindent=82.5px, linebackgroundcolor={\ifnum\value{lstnumber}=2 \color{red!15} \fi}]
# Negation: $\textbf{\color{OliveGreen}namespace}$ instead of $\textbf{\color{OliveGreen}not namespace}$
if $\textbf{\color{MidnightBlue}namespace}$:
    self.namespacesFilter = [ "prymatex", "user" ]
else:
    self.namespacesFilter = namespace.split()
\end{lstlisting}
\end{minipage}
 & 
 \begin{minipage}{\columnwidth}
A default \texttt{namespacesFilter} should be used if no \texttt{namespace} is given. However, the condition checks
the inverse.\\
 \textbf{Fix:} replace \texttt{namespace} in Line 2 by \texttt{not namespace}.
 \end{minipage}
 \\
\bottomrule
\end{tabular}
\end{table*}

\section{Background}
In this section, we introduce the necessary background for our approach. To begin with,
we start by describing the single token localization and repair task tackled by \realit~
and how previous techniques addressed this task by predicting token replacements
and learning from mutants. 

\subsection{Single token bug localization and repair}
In this work, we focus on the localization and repair of {\em single token bugs}. Single token bugs are bugs that can be repaired by replacing only a single program token (e.g. a variable or binary operator). For this reason, they are often easy to repair -- as only a single token has to be changed -- but hard to identify. Examples for single token bugs are given in Table~\ref{tab:ex}. 
Interestingly, single token bug localization and repair has previously only been addressed through training with mutants~\cite{pradel2018deepbugs, hellendoorn2019global, vasic2019neural,allamanis2017learning, richter2022learning, allamanis2021self}. Nevertheless,
real bug fixes for single token bugs -- which can be employed for training or testing -- are available in bug fix collections such as ManySStuBs4J~\cite{ManySStuBs4J} or TSSB-3M~\cite{richter2022tssb}.

\inlineheadingbf{Task description} Throughout this work, we view source code as a sequence of tokens $\mathcal{T} = t_0, t_1, t_2,\dots, t_n$. 
A single token bug can then be fixed by replacing a {\em single} token $t_l$ with another token $r$ in the same scope ($r = t_{l^\prime}$) or
coming from an external vocabulary ($r \in V$). 
To effectively localize and repair a single token bug, the following three tasks have to be performed: (1) the program $\mathcal{T}$ has to be {\em classified} to contain a bug, (2) the bug location $t_l$ has to be {\em localized} and then (3) the correct {\em repair} $r$
has to be identified. In practice, these three tasks are often modeled as {\em token replacement} operations. Let $\mathcal{T}$ be a program containing a single token bug and $\mathcal{T}^\prime$ be the corrected bug-free version, then the localization and repair model
is trained to perform the following operations:
\vspace{-1cm}
\begin{multicols}{2}
  \begin{equation}
   	\mathcal{T} \xrightarrow{\text{replace} (t_l,  r)} \mathcal{T}^\prime
  \end{equation}\break
  \begin{equation}
    \mathcal{T}^\prime \xrightarrow{\text{noop}()} \mathcal{T}^\prime
  \end{equation}
\end{multicols}
Here, we fix the buggy program $\mathcal{T}$ by replacing $t_l$ with $r$ and therefore translating it into $\mathcal{T}^\prime$.
Since $\mathcal{T}^\prime$ is {\em bug-free}, a change is not required ({\em noop}).

In practice, we train models to estimate the likelihood of each {\em possible} token replacement and select the most likely replacement to fix $\mathcal{T}$.

\subsection{Mutation}
Motivated by the general absence of real bug fixes at a sufficient scale,
 previous learning based localization and repair approaches~\cite{hellendoorn2019global, vasic2019neural, richter2022learning, allamanis2021self} mainly focused on training on {\em mutants}. Mutants are artificially generated (pseudo-)bugs that are introduced into a correct program via a mutation operator. For single token bugs, the mutation operator can be seen as a token replacement operator which can be inverted by a localization and repair model:
 \begin{equation}
   	\mathcal{T} \xrightarrow{\text{mutate} (t_l, r)} \mathcal{T}^\prime \xrightarrow{\text{replace} (t_l,  r^{-1})} \mathcal{T}
 \end{equation}
 For a dataset of bug-free programs (e.g. mined from open source projects), the mutation operator first introduces a token mutation by replacing a random token with a random other token. The token types are
 often specified (e.g. binary operators) such that the programs remains interpretable after the transformation. Afterwards, the localization and repair model is trained to invert the mutation process to obtain the original program.
 
While traditionally mutation operator are designed as a random process, previous work also tried to design more realistic mutation operators  by learning from real bug fixes~\cite{patra2021semantic}, by training an adversary to the repair model~\cite{allamanis2021self} or by finding replacements that naturally fit the context~\cite{richter2022learning}.

\subsection{Real bug fixes}
Real bug fixes are often obtained by scraping the commit history of public open source projects. During this process, commits are often classified as bug fixing based on certain keywords in the commit message~\cite{ManySStuBs4J}. Even though this process cannot guarantee that every collected commit is a real bug fix, it has been empirically shown~\cite{ManySStuBs4J} that the process is highly precise (e.g. over 90\% of all collected code changes were real bug fixes). In this work, we are interested in special types of {\em single token bug fixes}. Here, a bug is fixed by replacing only a single token: 
 \begin{equation}
   	\mathcal{T}_i  \xrightarrow{\text{replace} (t_l,  r)} \mathcal{T}_{i + 1}
 \end{equation}
Note that a (bug fixing) commit only represents a snapshot of the project at time $i$. Therefore, while it is highly likely $\mathcal{T}_i$ contains a single token bug which can be fixed by $\text{replace} (t_l,  r)$, we cannot guarantee that the bug fix is complete and $\mathcal{T}_{i + 1}$ is bug-free. 
\section{Methodology}
In this section, we introduce 
{\em \realit}~ as an effective training technique for bug localization and repair with Transformers. We start by giving a general overview of the training process for transferring and improving the performance of a localization and repair model trained solely on mutants.
Afterwards, we discuss the Transformer-based architecture used during training and the inference strategy we apply for localizing and repairing single token bugs in more detail.

\begin{figure*}[t]
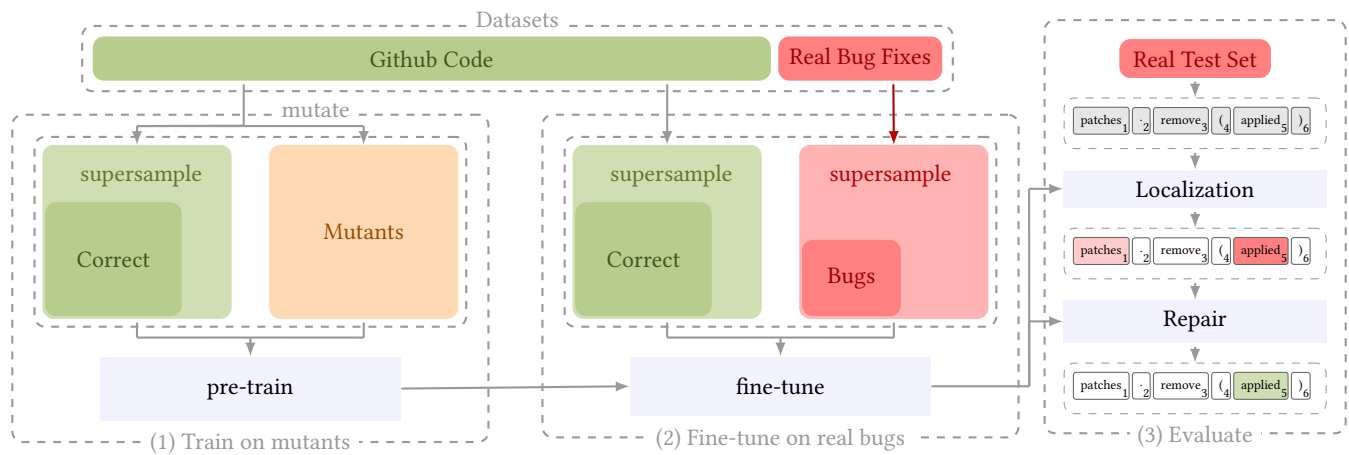

\begin{tikzpicture}

\node[draw=white!40!correct-green, fill=white!40!correct-green, rounded corners, inner sep=5pt, minimum width=9cm](correct-data){\textcolor{black!50!correct-green}{Github Code}};
\node[draw=white!50!red, fill=white!50!red, rounded corners, inner sep=4pt, right=0.1cm of correct-data](bug-data){\textcolor{black!40!red}{Real Bug Fixes}};
\node[dashed, draw=white!60!black, thick,rounded corners, inner sep=4pt, fit=(correct-data)(bug-data)](data-box){};
\node[above=-0.15cm of data-box, fill=white]{\textcolor{white!60!black}{Datasets}};

\node[draw=white!60!correct-green, fill=white!60!correct-green, rounded corners, below left=0.7cm and -2cm of data-box, minimum width=2.5cm, minimum height=2.3cm](correct-pretrain){};
\node[draw=white!40!correct-green, fill=white!40!correct-green, rounded corners, below left=-1.55cm and -1.85cm of correct-pretrain, minimum width=1.8cm, minimum height=1.5cm](correct-pretrain-inner){\textcolor{black!50!correct-green}{Correct}};
\node[above left=0.1cm and -2.2cm of correct-pretrain-inner]{\textcolor{black!30!correct-green}{supersample}};

\node[draw=white!70!orange, fill=white!70!orange, rounded corners, right=0.5cm of correct-pretrain, minimum width=2.5cm, minimum height=2.3cm](mutants){\textcolor{black!40!orange}{Mutants}};

\node[dashed, draw=white!60!black, thick,rounded corners, inner sep=3pt, fit=(correct-pretrain)(mutants)](pretrain-data){};

\node[fill=white!95!blue, below=0.38cm of pretrain-data, inner sep=8pt, minimum width=4cm](pretrain){pre-train};
\node[dashed, draw=white!60!black, thick,rounded corners, inner sep=8pt, fit=(pretrain-data)(pretrain)](pretrain-box){};
\node[below=-0.25cm of pretrain-box, fill=white]{\textcolor{white!60!black}{(1) Train on mutants}};

\draw ($(correct-data.south) + (-2.5, -0.05cm)$) edge[thick, draw=white!60!black] ($(correct-data.south) + (-2.5, -0.6cm)$)
		($(correct-data.south) + (-2.5, -0.6cm)$) edge[thick, draw=white!60!black] ($(correct-pretrain.north) + (0, 0.25cm)$)
		($(correct-pretrain.north) + (0, 0.25cm)$) edge[thick, draw=white!60!black,-latex] ($(correct-pretrain.north)$) 
		($(correct-pretrain.north) + (0, 0.25cm)$) edge[thick, draw=white!60!black] node[above right=0cm and 0.3cm, fill=white]{\textcolor{white!60!black}{mutate}}  ($(mutants.north) + (0, 0.25cm)$)
		($(mutants.north) + (0, 0.25cm)$) edge[thick, draw=white!60!black, -latex] ($(mutants.north)$)  ;
		
\draw ($(correct-pretrain.south) + (0, -0.05cm)$) edge[thick, draw=white!60!black] ($(correct-pretrain.south) + (0, -0.25cm)$) 
		 ($(correct-pretrain.south) + (0, -0.25cm)$) edge[thick, draw=white!60!black] ($(pretrain.north) + (0, 0.250cm)$) 
		 ($(pretrain.north) + (0, 0.250cm)$) edge[thick, draw=white!60!black, -latex] ($(pretrain.north)$) 
		 ($(mutants.south) + (0, -0.05cm)$) edge[thick, draw=white!60!black] ($(mutants.south) + (0, -0.25cm)$) 
		  ($(mutants.south) + (0, -0.25cm)$) edge[thick, draw=white!60!black] ($(pretrain.north) + (0, 0.250cm)$) ;


\node[draw=white!60!correct-green, fill=white!60!correct-green, rounded corners, right=1.4cm of pretrain-data, minimum width=2.5cm, minimum height=2.3cm](correct-train){};
\node[draw=white!40!correct-green, fill=white!40!correct-green, rounded corners, below left=-1.55cm and -1.85cm of correct-train, minimum width=1.8cm, minimum height=1.5cm](correct-train-inner){\textcolor{black!50!correct-green}{Correct}};
\node[above left=0.1cm and -2.2cm of correct-train-inner]{\textcolor{black!30!correct-green}{supersample}};

\node[draw=white!70!red, fill=white!70!red, rounded corners, right=0.5cm of correct-train, minimum width=2.5cm, minimum height=2.3cm](bugs-train){};
\node[draw=white!50!red, fill=white!50!red, rounded corners, below left=-1.05cm and -1.35cm of bugs-train, minimum width=1.3cm, minimum height=1cm](bugs-train-inner){\textcolor{black!40!red}{Bugs}};
\node[above left=0.6cm and -2.1cm of bugs-train-inner]{\textcolor{black!30!red}{supersample}};

\node[dashed, draw=white!60!black, thick, rounded corners, inner sep=3pt, fit=(correct-train)(bugs-train)](train-data){};

\node[fill=white!95!blue, below=0.38cm of train-data, inner sep=8pt, minimum width=4cm](train){fine-tune};
\node[dashed, draw=white!60!black, thick, rounded corners, inner sep=8pt, fit=(train-data)(train)](train-box){};
\node[below=-0.25cm of train-box, fill=white]{\textcolor{white!60!black}{(2) Fine-tune on real bugs}};

\draw ($(correct-train.south) + (0, -0.05cm)$) edge[thick, draw=white!60!black] ($(correct-train.south) + (0, -0.25cm)$) 
		 ($(correct-train.south) + (0, -0.25cm)$) edge[thick, draw=white!60!black] ($(train.north) + (0, 0.250cm)$) 
		 ($(train.north) + (0, 0.250cm)$) edge[thick, draw=white!60!black, -latex] ($(train.north)$) 
		 ($(bugs-train.south) + (0, -0.05cm)$) edge[thick, draw=white!60!black] ($(bugs-train.south) + (0, -0.25cm)$) 
		  ($(bugs-train.south) + (0, -0.25cm)$) edge[thick, draw=white!60!black] ($(train.north) + (0, 0.250cm)$) ;
		  
\draw ($(pretrain.east)$) edge[thick, draw=white!60!black, -latex] ($(train.west)$) ;

\draw ($(correct-data.south) + (3.13, -0.05cm)$) edge[thick, draw=white!60!black, -latex] ($(correct-train.north)$) ;
		
\draw ($(bug-data.south) + (0.44, -0.05cm)$) edge[thick, draw=black!30!red, -latex] ($(bugs-train.north) + (0, 0)$);


\node[draw=white!50!red, fill=white!50!red, rounded corners, inner sep=4pt, right=2.2cm of data-box, minimum width=2cm](eval-data){\textcolor{black!40!red}{Real Test Set}};

\node[below=0.27cm of eval-data](program-tokens){\input{figures/overview/program-tokens}};
\node[dashed, draw=white!60!black, rounded corners, inner sep=0.1pt, fit=(program-tokens)](token-box){};
\draw ($(eval-data.south) + (0, -0.05cm)$) edge[thick, draw=white!60!black, -latex] ($(token-box.north)$) ;

\node[fill=white!95!blue, below=0.27cm of token-box, inner sep=4pt, minimum width=3.5cm](loc){Localization};
\draw ($(token-box.south) + (0, -0.05cm)$) edge[thick, draw=white!60!black, -latex] ($(loc.north)$) ;

\node[below=0.27cm of loc](loc-tokens){\input{figures/overview/loc-tokens}};
\node[dashed, draw=white!60!black, rounded corners, inner sep=0.1pt, fit=(loc-tokens)](loc-box){};
\draw ($(loc.south) + (0, -0.05cm)$) edge[thick, draw=white!60!black, -latex] ($(loc-box.north)$) ;

\node[fill=white!95!blue, below=0.27cm of loc-box, inner sep=4pt, minimum width=3.5cm](repair){Repair};
\draw ($(loc-box.south) + (0, -0.05cm)$) edge[thick, draw=white!60!black, -latex] ($(repair.north)$) ;

\node[below=0.27cm of repair](rep-tokens){\input{figures/overview/rep-tokens}};
\node[dashed, draw=white!60!black, rounded corners, inner sep=0.1pt, fit=(rep-tokens)](rep-box){};
\draw ($(repair.south) + (0, -0.05cm)$) edge[thick, draw=white!60!black, -latex] ($(rep-box.north)$) ;

\node[dashed, draw=white!60!black, thick, rounded corners, inner sep=7.2pt, fit=(eval-data)(rep-box)](eval-data-box){};
\node[below=-0.24cm of eval-data-box, fill=white]{\textcolor{white!60!black}{(3) Evaluate}};

\draw ($(train.east) + (0, 0)$) edge[thick, draw=white!60!black] ($(train.east) + (1.3cm, 0cm)$) 
	($(train.east) + (1.3cm, 0cm)$) edge[thick, draw=white!60!black] ($(repair.west) + (-0.45cm, 0cm)$)
	($(repair.west) + (-0.45cm, 0cm)$)edge[thick, draw=white!60!black, -latex] ($(repair.west) + (0cm, 0cm)$) 
	($(train.east) + (1.3cm, 0cm)$) edge[thick, draw=white!60!black] ($(loc.west) + (-0.45cm, 0cm)$)
	($(loc.west) + (-0.45cm, 0cm)$)edge[thick, draw=white!60!black, -latex] ($(loc.west) + (0cm, 0cm)$) ;

\end{tikzpicture}
\caption{Overview over the \realit~ training and evaluation process}
\label{fig:overview}
\end{figure*}
\subsection{\realit: Training on mutants and real bugs}
To facilitate both mutants and real bug fixes, we design \realit~ as a {\em pre-train-and-fine-tune} 
approach. Therefore, we perform the training in two phases (as shown in  Figure \ref{fig:overview}).
In the first {\em pre-training phase}, we train \realit~ on artificially generated mutants introduced into  source code obtained by mining public open source repositories. Afterwards, in the second {\em fine-tuning} phase, we employ the pre-trained version of \realit~ to further train it on real bug fixes.

\inlineheadingbf{Pre-training with code mutants} During pre-training, we train our model similar to the way current localization and repair models are trained~\cite{hellendoorn2019global}. Here, the training objective is not to identify real bugs but rather to identify and transform mutated code snippets back into the real code snippets. For this task, we naturally start with a general corpus of {\em Github code} snippets (e.g. function implementations). This corpus can often easily be obtained by mining the recent version of popular open source projects. Since bugs are scarce in open source projects, we can safely assume that most of the code snippets are {\em likely correct}.
For training, we  {\em mutate} each code snippet in our corpus at max $k$ times which produces a dataset of at max $k$ {\em unique mutants} per code snippet\footnote{The number of code rewrites applicable to a code snippet and hence the number of unique mutants per code snippet is limited by design and might be lower than $k$. We never introduce mutation duplicates.}.
During our experiments, we decided to employ an unusually large number of mutants per code snippet ($k$ = 100) since we observed that this improves the performance after fine-tuning. We employ the original code corpus as training examples of unmutated {\em correct} code. 
Based on the two datasets, \realit~ is then trained to (1) distinguish mutants from real code, (2) identify the mutated location (if any) and (3) find the original replaced token. Since the dataset of mutants is up to $k$ times larger than the set of correct code snippet (by construction), we additionally {\em supersample}\footnote{During training, mutants and correct programs are sampled at the same frequency. As the set of mutants is up to $k$ times larger, we uniformly (super-)sample correct programs multiple times to match the number of mutants seen during training.} each correct code snippet such that \realit~ is trained on correct and mutated code snippets at the same frequency. This avoids that the model is biased towards detecting mutants.

\inlineheadingbf{Learning to fix real bugs with bug fixes} In the second phase, we aim to further optimize the performance of \realit~ for localization and repair of real bugs by fine-tuning on real bug fixes. For the fine-tuning process, we adopt a pre-trained version of \realit~ obtained from the previous phase. Then, we continue the training now on realistic buggy and bug-free code. As examples for realistic buggy code, we employ the code related to real bug fix {\em before} the fix is applied. Bug-free code is again obtained by using the original Github corpus. 
During training, \realit~ is now fine-tuned to (1) distinguish {\em real buggy code} from bug-free code, (2) identify the bug location (if any) and (3) imitate the original bug fix. Since now, the code corpus is usually much larger than the set of bug fixes, we supersample the buggy programs to match the correct programs in their frequency.

We believe that pre-training and fine-tuning can have an orthogonal effect on the localization and repair model (which we aim to explore during our evaluation). Due to the mutation process, the pre-training phase is more tailored towards identifying correct
programs and deviations from them. In contrast, the fine-tuning phase aims to teach the model the difference between real correct programs and real buggy programs (and how they can be fixed). 

\subsection{Model architecture}
In the section, we discuss the neural model employed by \realit~
to learn localization and repair of single token bugs. 
Since our main focus is to study the effect of mutants and real bug fixes
on the training process, we employ a fairly standard Transformer-based model~\cite{hellendoorn2019global} for learning bug localization and repair.

\inlineheadingbf{Probabilistic model} For the task of single token localization and repair, programs are represented as a sequence of tokens $\mathcal{T} = t_0, t_1, t_2, \dots, t_n$ where each token $t_l$ represents a potential bug location. Single token bugs are fixed only by replacing single tokens $t_l$ with another token $r$ ($\text{replace} (t_l,  r)$). In the following, we model localization and repair as a joint probability distribution over all potential bug locations and repairs $\{ \langle l, r \rangle \mid t_l \in \mathcal{T} \cup \{\mathtt{NOOP}\} \text{ and } r \in \mathcal{T} \cup V \}$:
\begin{equation}
p ( \langle l, r \rangle \mid \mathcal{T}) = p_{\text{loc}} ( l \mid \mathcal{T}) \cdot p_{\text{repair}}( r \mid l, \mathcal{T})
\end{equation}
Here, localization and repair is factorized into first localizing a bug location ($p_{\text{loc}} ( l \mid \mathcal{T})$) and then finding a repair dependent on the bug location ($p_{\text{repair}}( r \mid l, \mathcal{T})$). For localization, we include a special \texttt{NOOP} location that indicates that $\mathcal{T}$ is bug-free and no repair is necessary. In practice, we implement the probability distributions similar to pointer networks~\cite{gu2016incorporating} (with the addition of an external vocabulary  for repair). 

\inlineheadingbf{Neural code representation} To learn a neural code representation, we learn
a neural encoding function $\textbf{e}(t_i)$ that maps each token $t_i$ to a vector representation.
For this, we employ a BPE subtoken encoder~\cite{sennrich-etal-2016-neural} (with a vocabulary of 10K subtokens) to obtain an initial token embedding
by averaging the embedding of its subtokens.
Afterwards, we encode the sequence of token embeddings via a Transformer encoder~\cite{devlin-etal-2019-bert} (with relative position encodings~\cite{shaw2018self}) to obtain a contextualized vector representation $\mathbf{r}_i = \textbf{e}(t_i)$. 

\inlineheadingbf{Localization \& Repair module}
To finally compute the probability distribution over bug locations and repairs, we employ individual
modules for localization and repair based on the computed vector representation $\mathbf{r}_i$.
The {\em localization module} is a multilayer perceptron that computes a bugginess score for each potential bug location
based on the vector representation $\mathbf{r}_i$ and the original token embedding.
The objective of the localization module is to learn how likely the token $t_i$ does not fit its surrounding context (represented by $\mathbf{r}_i$). The localization probability is computed as a softmax distribution over all potential bug locations.
The {\em repair module} is designed similar to CopyNet~\cite{gu2016incorporating}. Given the vector representation $\mathbf{r}_i$ of a potential bug location, the repair module computes a repairing score between the bug location
and each repair candidate at token $t_j$ (represented by $\mathbf{r}_j$). In addition, a similar score
is obtained based on token embeddings of an external vocabulary $V$ (e.g. other binary operators). The repair probability score is then computed as softmax distribution between all repair candidates.

\subsection{Finding and repairing real bugs}
After the successful training process, the localization and repair models are typically confronted with 
new unseen programs with the objective to identify a potential bug and repair it.
This is typically done~\cite{vasic2019neural} by finding the {\em most likely} repair for the {\em most likely} bug location (according to the model). 
However, the {\em most likely} repair at the {\em most likely} bug location might not always
be {\em meaningful}. For example, while the model might be confident that a bug is located at a certain
location, there might not be a suitable repair candidate that can actually fix the bug.
For this reason, we propose an alternative strategy: Instead of taking the most likely repair for the most likely bug location, we search for the most likely {\em meaningful} combination of bug location and its repair (and thus ignoring bug localizations that cannot be fixed by the model).

\inlineheadingbf{Beam search decoding} RealiT implements this search strategy via a beam search decoding~\cite{kalchbrenner2013recurrent}. Here, we iterate the top-$k$ bug locations according to the model and for each bug location we again search for the top-$k$ token repairs. During this process, we only store pairs that are assigned the highest likelihood {\em together}. Afterwards, we filter the candidate pairs to only {\em meaningful} repairs: If the model predicts a bug location then this should be fixed by a different token of the same type. Combinations of the special \texttt{NOOP} location and repair are always meaningful (since nothing will be changed).
Finally, as a result, RealiT computes the most likely meaningful repair operation:
\begin{equation}
 \text{replace} (t_{l^\prime},  r^\prime) \text{ with } \langle l^\prime, r^\prime \rangle = \text{argmax } p ( \langle l, r \rangle \mid \mathcal{T}) 
\end{equation}

\subsection{Implementation}
To effectively measure the impact of mutants and real bug fixes on the training process and to 
exclude variances due to implementation details, we implemented \realit~ together with several baselines considered during our evaluation in a unified framework\footnote{https://github.com/cedricrupb/nbfbaselines}. In particular, we followed the design of Hellendoorn et al.~\cite{hellendoorn2019global} by implementing a common localization and repair framework with exchangeable components (e.g. token encoder, localization and/or repair modules). In the process, we reimplemented or reused state-of-the-art components for all employed subcomponents. For example, \realit~ and all Transformer-based baselines are built upon
the official BERT implementation from the \texttt{transformer} library~\cite{wolf-etal-2020-transformers}. The localization and repair modules together with the graph-based baselines are implemented closely to the implementation of the PyBugsLab model~\cite{allamanis2021self}.  
In addition, we reimplemented the code preprocessing pipelines for tokenization~\cite{hellendoorn2019global} and graph construction~\cite{allamanis2021self} (used for the graph-based baselines) in independent libraries to facilitate reuse. 
Finally, we plan to release all trained checkpoints of \realit~ and all evaluated models. We think that these are not only valuable for reproducing our results but also provide easy access to effective models in neural bug localization and repair.

\section{Evaluation}
We evaluate \realit~ on localization and repair of single token bugs in Python. To guide our evaluation, we specifically designed individual experiments to address the following research questions:
\begin{description}
\item[RQ1] Can \realit~ improve the single token bug localization and repair performance in comparison to techniques purely trained on mutants?
\item[RQ2] Is pre-training on mutants necessary for achieving a high localization and repair performance?
\item[RQ3] Can training with mutants alone be sufficient for achieving a high performance?
\item[RQ4] Are real bug fixes still helpful if the number of real bug fixes available for training is further limited?
\end{description}
In RQ1, we compare \realit~ with various techniques purely trained on mutants. RQ2 and RQ3 are designed to explore the effect of mutation and real bug fixes on the training process. Especially, since real bugs are hard to obtain, we are interested whether they are really necessary for training effective bug localizer and repairer. 
Finally, in RQ4, we explore how many real bug fixes are necessary in practice to improve the localization and repair performance.

\subsection{Bug types}
To facilitate both mutants and real bug fixes, we require bug types that can be introduced by existing
mutation operators and where examples of real bug fixes are available. For this reason, we focus on the {\em four} single token bug types in Python that
can be generated by existing mutation operators~\cite{derezinska2014analysis} and for which we can obtain real bug fixes
for training and evaluation~\cite{richter2022tssb}. In the following, we describe the bug types together with the employed mutation operator in more detail.

\inlineheadingbf{Variable Misuse} As the main carrier of the program state, variables are abundant in source code. Therefore,
variable misuses easily occur when a developer accidentally uses the wrong variable name instead of the intended one. As specified
by Allamanis et al.~\cite{allamanis2017learning}, the usage of a wrong variable is considered as a variable misuse if the wrong usage refers to a {\em local} variable which can be fixed by replacing it with another {\em locally} defined variable.

{\em Mutator:} For generating variable misuses, the mutator replaces a usage of a locally defined variable with another random variable
defined in the same context. 

\inlineheadingbf{Wrong Binary Operator} As a traditional example for a mutation type in mutation testing~\cite{derezinska2014analysis}, 
wrong binary operator bugs appear when a binary operator is corrupted with a type-equivalent operator (e.g. \texttt{==} is replaced by \texttt{!=}, but not by \texttt{<}\texttt{<}).
For training \realit, we consider all types of binary operators including Boolean, arithmetic, comparison and bitvector operators.

{\em Mutator:} Wrong binary operator mutants are generated by replacing an binary operator with another random binary
operator of the same type.

\inlineheadingbf{Wrong Unary Operator} In addition to binary operators, we also consider two types of wrong unary operator bugs: logical
and arithmetic negation. In contrast to binary operators, wrong unary operators are often not replaced but primarily occur when a unary operator is missing 
or accidentally added. This includes for example a forgotten logical negation in a condition or an accidentally added arithmetic inversion
during a calculation.

{\em Mutator:} Wrong unary operator mutants are generated by randomly dropping or inserting
negation operators (either arithmetic or logical) in front of an identifier. To ensure that inserted negations are semantically
meaningful, negations are inserted dependent on the context (e.g. logical negations in conditions and arithmetic negations in arithmetic expressions).

\inlineheadingbf{Wrong Literal} Another common bug type produced by mutation operators are literal replacements. Similar to mutation testing~\cite{derezinska2014analysis}, we are limited
to literal replacements from {\em finite} sets of common literal types. This naturally includes Boolean literal replacement by replacing \texttt{True} with \texttt{False}
and vice versa, but also integer replacements from the set \texttt{-2},  \texttt{-1},  \texttt{0},  \texttt{1},  \texttt{2}.

{\em Mutator:} Mutants are generated by replacing literals from a set of predefined literals with another random literal
of the same set and type.

\subsection{Datasets}
For training and evaluating \realit, we require two types of datasets: a general {\em Github corpus} of code snippets and a dataset of {\em real bug fixes}. To achieve comparable results, we employ
existing datasets (if available). For the same reason, we also decided to focus on bugs in Python function implementations. 

\inlineheadingbf{Github code corpus} As a general corpus of Python code, we employ the ETH Py150k dataset~\cite{raychev2016probabilistic} containing over 150k program files from popular Python projects. The dataset is split into 100k files for training and 50k files for testing. During our evaluation, we employ the same split and, hence, train only on Python functions obtained from train split. The test split is used for evaluating the performance on mutants. We extract all top level functions and deduplicate the datasets such that the Python functions used for training only occur once in the training dataset and do not occur in the test split. In total, our training corpus contains more than 360k Python function implementations (after filtering). 

\inlineheadingbf{Real bug fixes} For obtaining real bug fixes at a sufficient scale, we employ
the SSB-9M dataset~\cite{richter2022tssb} of over 9M general single statement bug fixes in Python. The dataset
does not include the necessary implementation code itself but references the original commits in the repositories in addition to other useful metadata. This includes information about the code change as a Unix diff and whether the code change appears inside a function. Based on this information, we first pre-filtered the dataset for bug fixes that likely fall into one of our bug categories. After the filtering process, we then mined the function code from the original repositories. Since not all bug types can be identified purely on the code difference (e.g. a variable misuse requires that all variables are defined in scope), we filtered and deduplicated the resulting dataset of buggy Python functions for a second time. This process has lead to around 35k examples of real bug fixes that match at least one of our bug types.
Finally, we use 33k examples for training and hold out around 2k examples as a validation set used during training.

\inlineheadingbf{Test benchmarks} We employ two test benchmarks to evaluate the 
performance on the localization and repair task. To evaluate the localization and repair performance
on real bugs, we employ the PyPIBugs benchmark~\cite{allamanis2021self}. The benchmark is a dataset of 2374 real-world single statement bugs and their fix derived from open source projects. The benchmark is hand-filtered and therefore it is likely that each included bug represents a real world bug. We only considered single token bugs (which excludes argument swaps) in functions where the implementation is still publicly accessible\footnote{The benchmark consists of references to bug fixing commits. We found that not all bug fixing commits were publicly accessible at the time of writing.}. This produced a real world test benchmark of 2028 real-world bugs. To avoid an overlap between train and test set, we excluded all near duplicates~\cite{allamanis2019adverse} from our training datasets. Additionally, we also employ the test portion of the Github corpus as a mutant benchmark. For this, we extend the corpus of correct code snippets with up to 9 mutants per snippet.


\begin{table*}[t]
  \caption{Evaluation results for bug detection and repair on mutants and real bug fixes}
  \label{tab:results}
  \centering
  \begin{tabular}{l c  c c c c c c c c}
    \toprule
    && \multirow{2}{*}{\shortstack[c]{FPR\%}} & \multicolumn{3}{c}{\bfseries Real Bugs (PyPIBugs)} && \multicolumn{3}{c}{\bfseries Mutants} \\
    \cmidrule{4-6} \cmidrule{8-10}
   &&& Joint & Loc. & Repair && Joint & Loc. & Repair \\
    \midrule
    RNN~\cite{vasic2019neural} && 33.92 & 9.47 & 13.36 & 47.88 && 52.39 & 61.49& 80.06 \\
    Transformer~\cite{hellendoorn2019global} && 25.59 & 18.98 & 23.02 & 59.52 &&  74.04 & 81.26 & 88.74  \\
    GNN~\cite{allamanis2021self} &&25.29 & 18.24 & 23.82 & 53.74 && 66.11 & 75.08 & 84.59 \\
    GREAT~\cite{hellendoorn2019global} &&29.98 & 19.03 & 23.62 & 56.31 && 70.76 & 78.84 & 86.99  \\
    \realit~(ours) && 29.53 &  \textbf{39.00} & \textbf{44.23} & \textbf{73.52} && 67.09 & 75.95 & 85.66  \\
    \midrule
        \realit~ - without beam search decoding && \underline{22.96} & \underline{36.69} & \underline{41.86} & \textbf{73.52} && 65.27 & 74.08 & 85.66  \\
    \realit~ - without pre-training on mutants &&\textbf{19.32} & 12.67 & 16.07 & 40.38 && 2.37 & 8.58 & 31.34  \\
    \realit~ - without fine-tuning on real bug fixes &&33.41 &25.10 & 30.92 & 65.53 && \underline{77.41} & \underline{84.49} & \underline{90.18}  \\
     \realit~ - with fine-tuning on postfix mutants && 27.72 & 27.66 & 32.64 & 67.40 && \textbf{78.95} & \textbf{85.12} & \textbf{90.55}   \\
     \realit~ - with reduced mutant frequency (5x) &&31.75 & 33.28 & 38.56 & 68.59 && 65.32 & 74.74 & 83.90  \\
    
    \bottomrule
  \end{tabular}
\end{table*}

\section{Result}
In this section, we discuss our evaluation results with the ultimate goal of answering our research questions.

 \subsection{RQ1: \realit~ in comparison?}
For answering the first research question, we evaluate whether \realit~ improves the single token bug localization 
and repair performance by training with real bug fixes.
Since we are interested in the impact of real bug fixes on the training process, we compare our results with several baseline algorithms trained purely on mutants. For the comparison, we consider bug localization and repair models
based on {\em recursive neural networks} (RNN)~\cite{vasic2019neural},  {\em transformers} (absolute positions)~\cite{hellendoorn2019global},  {\em graph neural networks} (GNN%
\footnote{Our evaluation setup differs slightly from \cite{allamanis2021self} in that only individual function implementations are considered. Therefore, graph level information that would require access to the implementation context cannot be computed. }%
)~\cite{allamanis2021self} and {\em GREAT}~\cite{hellendoorn2019global}. 
All baseline models are trained in a supervised setting purely on mutants. The training dataset is constructed similar to the pre-training dataset used for \realit~ (with $k$ = 5 mutants injected).
The baselines are trained for 300 epochs (a 200k examples per epoch) with early-stopping on our validation set. 
For \realit, we skip the last epoch and instead fine-tune on real bug fixes. 
 
 \inlineheadingbf{Real-world performance} To begin with, we start by considering
 the performance of \realit~ on our real-world benchmark. Table~\ref{tab:results} provides
 an overview over our evaluation results. We measured the {\em joint} accuracy of localizing
 and repairing real bugs (Joint) in addition to the {\em localization accuracy} (Loc.) of finding the bug location and
 the {\em repair accuracy} (Repair) of finding the real bug fix given the bug location. In this section,
we focus on the upper part of the table and we consider the results on our real-world benchmark.
 
 We observe that \realit~ significantly outperforms all baseline algorithms trained purely on mutants 
 both in localization and repair of real single token bugs. Interestingly enough, we find that the highest relative
 gain obtained from fine-tuning on real bug fixes can be achieved for the localization performance (with a nearly 2x improvement).
 This indicates that for effective localization of human made bugs we actually need to learn from human made bug fixes.
Still, the bug localization remains harder than bug repair as \realit~ can fix more than 73\% of all bugs when the bug location is given. Therefore, it is important to investigate into better strategies for bug localization (potentially by integrating techniques from static analysis).
 
Finally, the fact that significant performance improvements are observable in both localization and repair suggests that real bug fixes exhibit exploitable statistics for localization and repair.  We will further explore this in RQ2 and RQ3.
 
 \inlineheadingbf{Localization and repair of mutants} While our ultimate goal is to find and fix real bugs, we also
 measured the localization and repair accuracy of \realit~ for artificial mutants. Surprisingly, we observe that \realit~
 performs worse than most baseline models both in localization and repair after fine-tuning on real bug fixes. Interestingly,
 this is {\em not} a limitation of the \realit~ model as the version of \realit~ trained purely on mutants performs competitively
 or even better than all baselines in localizing and repairing mutants. Therefore, the fine-tuning on real bugs encourages
 \realit~ to ``forget'' some (potentially spurious) patterns that were used to detect mutants but do not help for identifying
 real bugs. In addition, this provides further evidence that there might exist mutants that either do not represent real bugs
 or represent bugs that are highly unlikely to appear in reality. Finally, this observation is also interesting 
 for the evaluation of localization and repair models. As there is clearly no correlation between the real world performance and
 the performance on mutants when the model is fine-tuned on real bug fixes, performance gains on mutants independent from
 real world performance become difficult to interpret.

\inlineheadingbf{False positives} We also measured the false positive rate (FPR) of \realit~ on bug-free code snippets.
 Here, we employ the original test set of our Github corpus. Our results are also summarized in Table~\ref{tab:results}. We observe
 that \realit~ has a false positive rate comparable to the other baselines -- only outperformed by the Transformer and GNN. However, we believe that an increase of 3\% more false positives is still acceptable as \realit~ localizes and fixes nearly twice as many real bugs. In addition, we also evaluate a version of \realit~ without {\em beam search decoding} (i.e. using the repair with the highest likelihood). The results are also shown in the lower part of Table \ref{tab:results}. We observe that while beam search decoding improves the localization performance by up to 3\%, it also induces a worse false positive rate compared to the model without beam search decoding. This is a common trade off between a higher true localization performance with a worse false positive rate.

 \subsection{RQ2: Are mutants necessary?}
 As we have seen in RQ1, fine-tuning on real bug fixes does improve localization and repair of real bugs. This raises
 the question whether mutants are necessary for the performance gain or if the same performance can be achieved with real bugs
 alone. To answer this question and therefore RQ2, we trained two additional versions of \realit: (1) a version of \realit~ that is not pre-trained on mutants and (2) a version of \realit~ that is not fine-tuned on real bug fixes. We evaluate both versions again on all benchmarks.
 
\inlineheadingbf{Mutants vs real bugs} We start by comparing the two new versions of RealiT. Our evaluation
results are summarized with all other results in Table~\ref{tab:results}. We observe that training on mutants
outperforms training on real bug fixes only. It is likely that \realit~ overfits the {\em smaller} training dataset of real bug fixes
and therefore fails to generalize to complete new unseen bugs. In contrast, the version purely trained on mutants
has learned from a variety of mutants during training (some of which are likely similar to real bugs). However, when evaluated on bug-free code snippets only, we see that \realit~ trained only on real bug fixes clearly outperforms all other techniques in terms of 
false positive rate. This could again indicate that some mutants in the training dataset are not bug inducing (e.g. a mutation that replaces \texttt{<=} with \texttt{!=} without changing the function behavior) which guides the model to detect these structures in bug-free code.

\inlineheadingbf{Training with mutants and real bugs} We now compare the two variants of \realit~ with our original
\realit~ model. We find that fine-tuning on real bug fixes significantly improves the performance of \realit~ over the already
strong baseline of training on mutants alone. Interestingly enough, this does not only hold for localization and repair of real bug fixes
but also on the false positive rate on bug-free code. This shows that pre-training on mutants and fine-tuning on real bugs combines the strengths of both the high localization and repair performance (by training on mutants) and the bug detection accuracy (by training on real bugs). Therefore, we see that mutants are {\em necessary} to achieve the high performance of \realit~ but fine-tuning on real bugs provides additional improvements.

\inlineheadingbf{Effect on individual bug types} Since the effect of pre-training and fine-tuning seems to be complementary, we are also interested in how the training affects the performance on individual bug types. Table~\ref{tab:cat-results} summarizes our results on the real-world test benchmark divided into single token bug types. First of all, we again find that training on both mutants and real bugs does improve performance in both localization and repair on all bug types. However, the margin of the performance gain is dependent on the bug type. For example, we see the highest improvement for \texttt{Wrong Binary Op} where training on real bugs alone already yields high performance. To answer our research question also for individual bug types, pre-training on mutants can also be {\em crucial} for the performance on individual bug types (where we observe a significant improvement for at least three bug types \texttt{Wrong Assign Op}, \texttt{Wrong Literal} and \texttt{Variable Misuse}).

\begin{table*}
  \caption{Evaluation results for bug detection and repair on different bug types}
  \label{tab:cat-results}
  \centering
  \begin{tabular}{l  c c c c c c c c c c c c}
    \toprule
    \multirow{2}{*}{Bug type} && \multicolumn{3}{c}{\bfseries \realit} && \multicolumn{3}{c}{\bfseries Mutants only} && \multicolumn{3}{c}{\bfseries No Mutants} \\
    \cmidrule{3-5} \cmidrule{7-9}\cmidrule{11-13}
   && Joint & Loc. & Repair && Joint & Loc. & Repair && Joint & Loc. & Repair\\
    \midrule
    Wrong Assign Op && \textbf{20.45} & \textbf{29.54} & \textbf{70.45} && \underline{9.10} & \underline{13.64} & 52.27 && 2.27 & 2.27 & \underline{65.91} \\
    Wrong Binary Op &&  \textbf{56.34} &  \textbf{59.15} &  \textbf{84.51} && 14.08 & 28.17 & 39.44 && \underline{30.99} & \underline{35.21} & \underline{70.42} \\
    Wrong Boolean Op &&  \textbf{42.31} &  \textbf{42.31} &  \textbf{95.05} && \underline{23.08} & \underline{24.18} & \underline{93.41} && 21.43 & 21.97 & 81.87 \\
    Wrong Comparison Op &&  \textbf{36.95} &  \textbf{51.47} &  \textbf{67.00} && 19.70 & \underline{35.22} & \underline{57.6}4 && \underline{23.40} & 33.74 & 57.14 \\
    Wrong Literal &&  \textbf{24.42} &  \textbf{32.56} & \underline{76.74} &&  \underline{19.77} &  \underline{22.09} & \textbf{77.91} && 9.30 & 12.79 & 46.51 \\
    Variable Misuse &&  \textbf{39.87} &  \textbf{42.62} &  \textbf{71.75} &&  \underline{28.73} &  \underline{31.88} &  \underline{65.13} && 7.43 & 9.04 & 25.75 \\
    \bottomrule
  \end{tabular}
\end{table*}

 \subsection{RQ3: Are mutants sufficient?}
Our evaluation for RQ2 has shown that training on mutants is crucial for obtaining high performing \realit~ models. Still, it is not clear whether mutants on its own can be {\em sufficient} for training \realit. In other words, there might exist a mutation configuration that achieves the same performance trained on mutants based on the same base datasets. To answer RQ3, we designed several experiments.

\begin{figure}[t]
	\begin{subfigure}[t]{1\linewidth}
		\begin{adjustbox}{max width=1\linewidth}
			\includegraphics{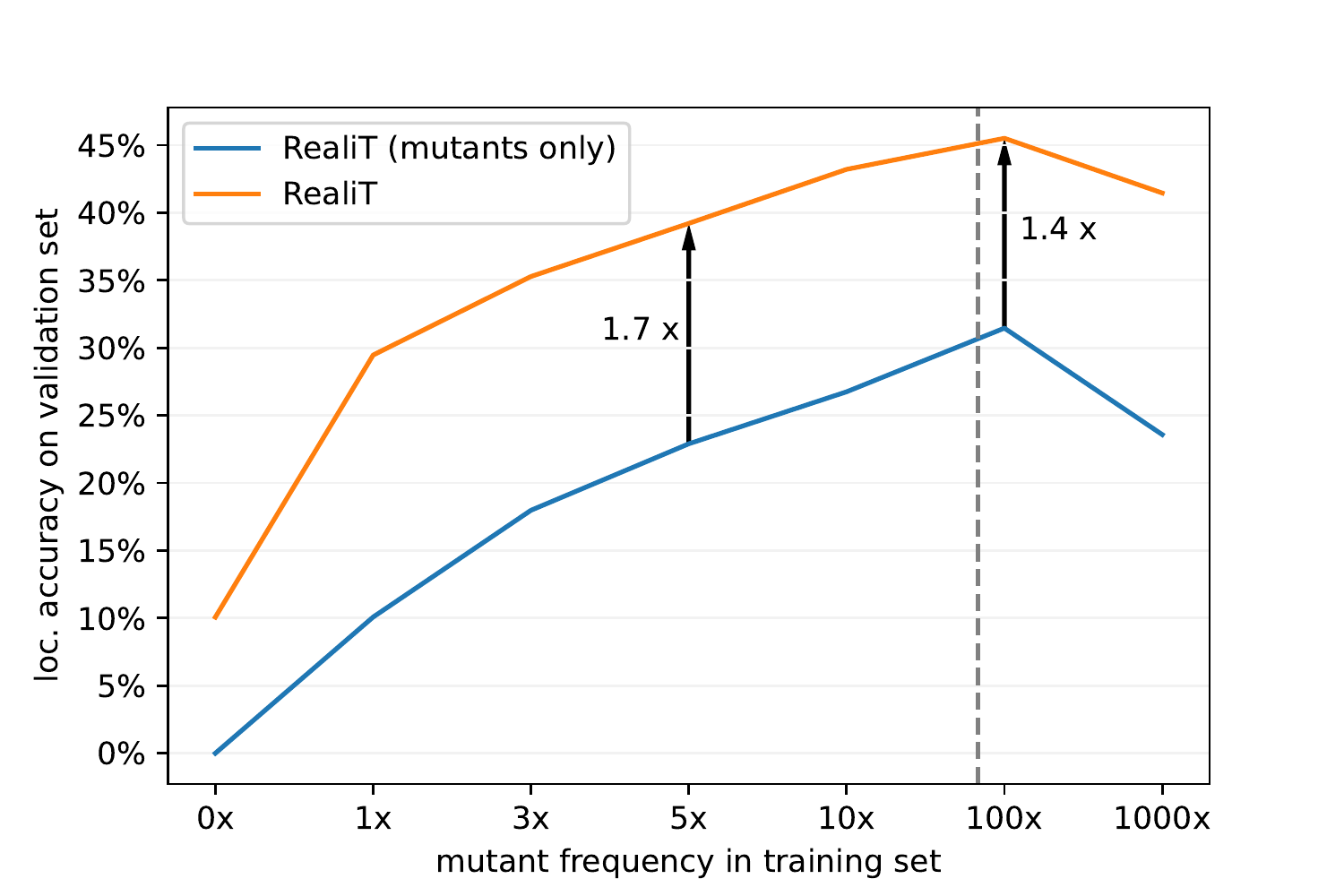}
		\end{adjustbox}

		\caption{Localization accuracy \label{fig:mutantloc}}
	\end{subfigure} 
	\begin{subfigure}[t]{1\linewidth}
		\begin{adjustbox}{max width=1\linewidth}
			\includegraphics{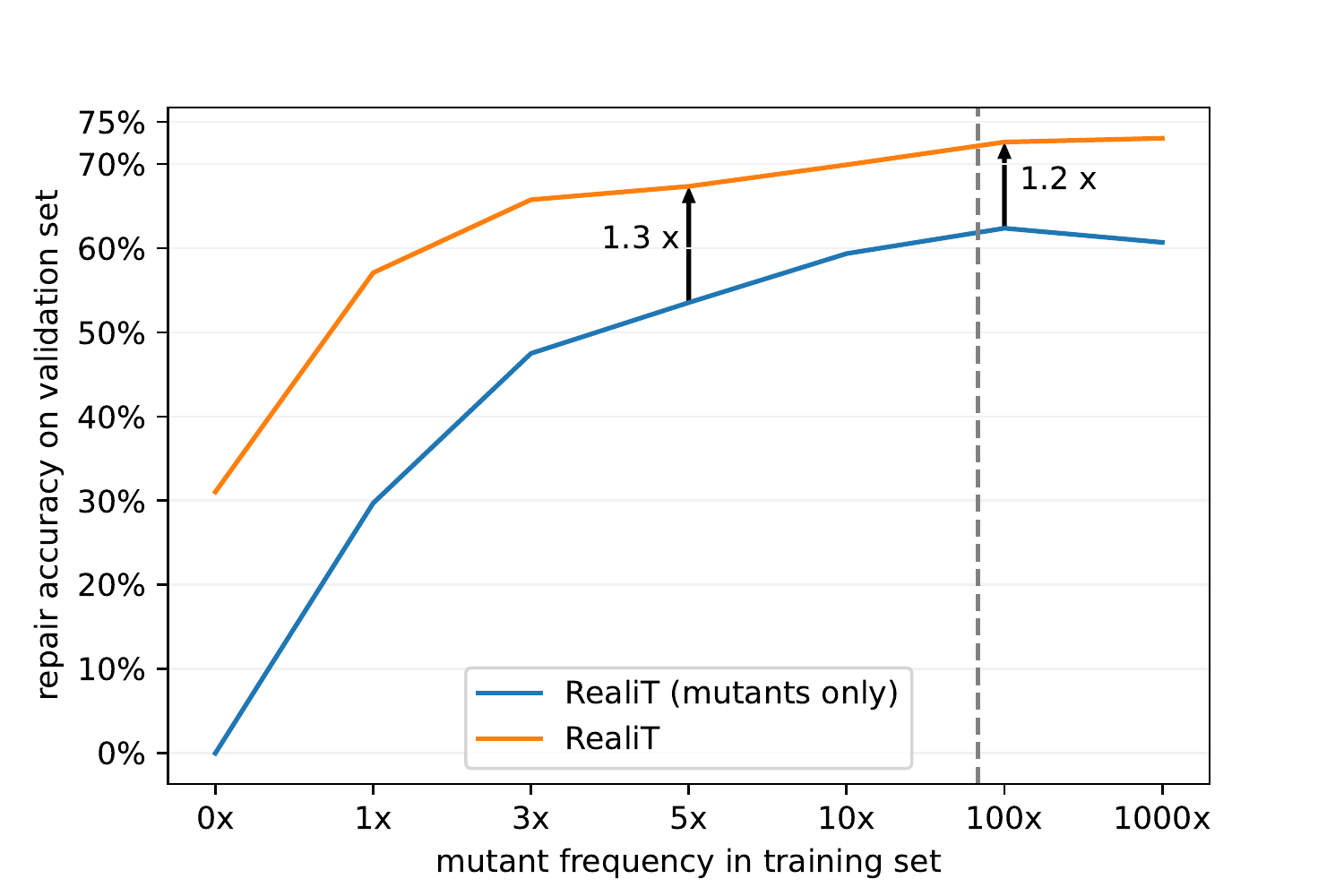}
		\end{adjustbox}

		\caption{Repair accuracy \label{fig:mutantrep}}
	\end{subfigure} \hfill
	\caption{Effect of mutant frequency during training on the real world validation performance. The gray dotted line represents the average number of unique mutants that can be generated per code snippet.}\label{fig:mutant-loc-repair}
\end{figure}

\inlineheadingbf{Mutation frequency} We trained several versions of \realit~
by varying the mutation frequency (up to 1x, 3x, 5x, 10x, 100x and 1000x unique mutants per code snippet). For the comparison, we measured the performance of each trained model before and after fine-tuning on real bug fixes. The models are evaluated on our real bugs validation set. Figure \ref{fig:mutant-loc-repair} gives an overview of our results for bug localization and repair accuracy independently. The configuration 0x represents a version of \realit~ only trained on real bug fixes.
First of all, in contrast to common believe~\cite{hellendoorn2019global}, we observe that increasing the number of mutants up to 100x generated mutants per code snippet leads to a performance improvement for both localization and repair\footnote{We observe the same trend for joint localization and repair which is not shown here for brevity.}. This is surprising as the number of unique mutants per code snippet is limited (with an average of 85 unique mutants per code snippet) and, henceforth, buggy programs with more mutant candidates are oversampled. Still, we found that increasing the limit of mutants beyond 100x (and thereby oversampling code snippets in our dataset that provide up to 200k unique mutants) actually decreases the localization and repair performance. 

Now, when also considering the performance on the validation set after fine-tuning, we find that fine-tuning always provides a significant boost over the models trained solely on mutants for both localization and repair accuracy. However, we still observe that the performance gain for localization is higher than
for repair (especially as we increase the number of mutants). Surprisingly, we also observe that the gap
between the model performances before and after fine-tuning on real bug fixes shrinks as we increase the number of mutants generated per code snippet (up to 100x). While this could indicate that simply scaling the number of mutants
can be sufficient for achieving a high repair accuracy, the gap actually starts increasing again after scaling beyond 100x mutants per code snippet. 
Therefore, we can conclude that while mutants alone can significantly improve the performance they are {\em not sufficient} in our training setup for achieving the same high performing localization and repair models as obtained by fine-tuning on real bug fixes with \realit.
 
 \inlineheadingbf{Training with postfix mutants} 
 While it seems that mutants alone introduced in \textit{arbitrary} code from our code corpus is not sufficient for closing the performance graph to fine-tuned models, it is unclear whether mutants can be sufficient when introduced in an implementation context that is more \textit{typical} for a real bug. 
  To test this,
 we designed an additional experiment where we fine-tuned \realit~ on {\em postfix mutants} (i.e. mutants that are produced by first applying
 the real bug fix and then reintroducing a bug with a mutation operator). 
Our results are also shown in Table~\ref{tab:results}. We observe that even fine-tuning on postfix mutants provides a slight boost in performance both on localization and repair of mutants and real bugs. Surprisingly, the boost is slightly higher for real bugs than for mutants even though we only increased the number of mutants. Still, we find that a model trained on real bug fixes clearly outperforms a model trained on postfix mutants when evaluated on real bug fixes. Since the performance of detecting mutants does not decrease when training on postfix mutants (similar to scaling the number of mutants generated), we actually conclude that the performance gain is likely a scaling effect and can not necessarily be attributed to the mutation context. 

In total, we find that training on mutants alone is not sufficient for achieving
a high performing \realit~ model. In addition, our results show that training on real bug fixes is especially helpful for the localization of real bugs which is hard to obtain by training on mutants alone. 

\subsection{RQ4: How many bug fixes are necessary?}
\begin{figure}[t]
		\begin{adjustbox}{max width=1\linewidth}
			\includegraphics{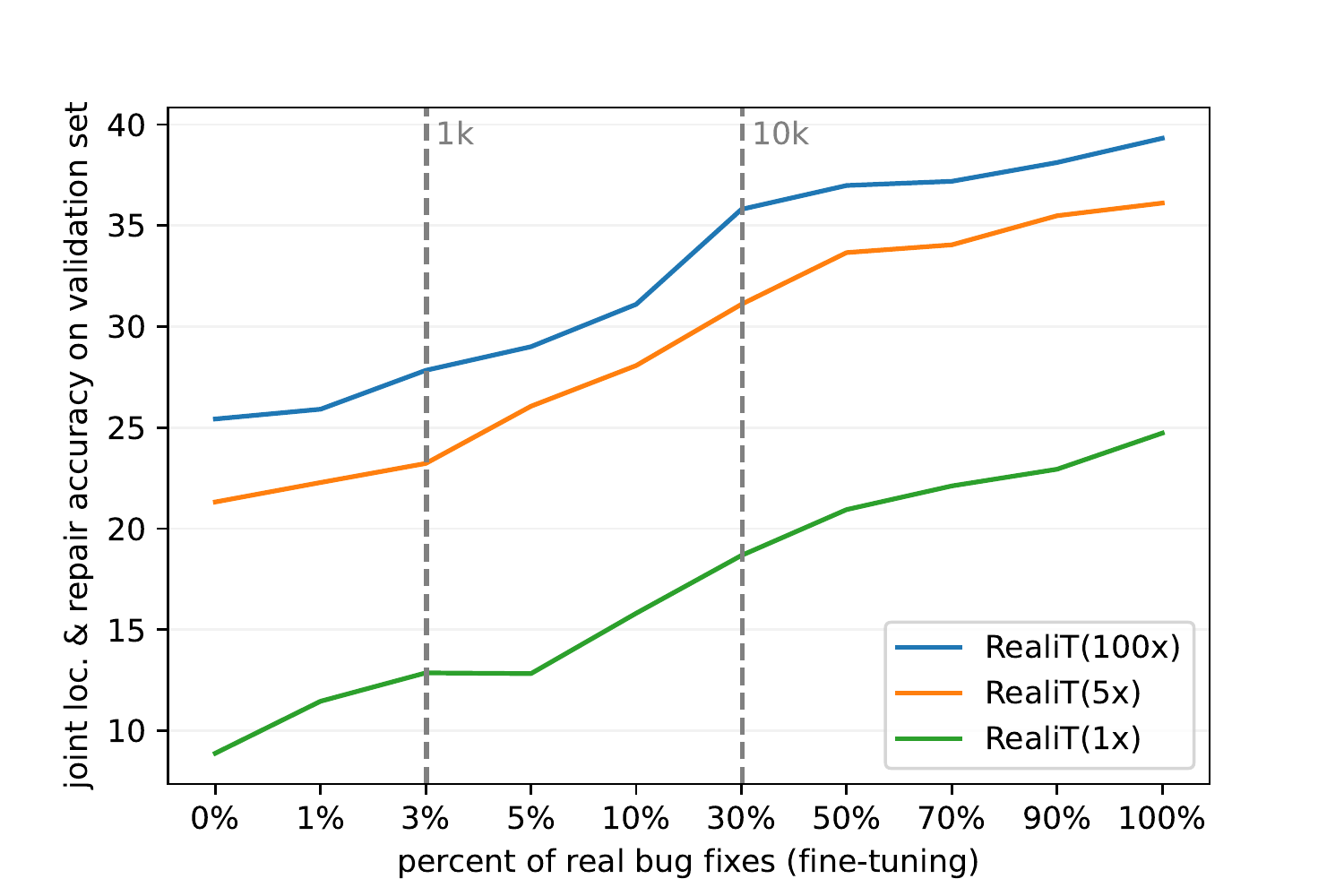}
		\end{adjustbox}
	\caption{Effect of real bug fixes on the fine-tuning performance on the validation set. The x-axis is the percentage of the bug fix dataset used for fine-tuning. Gray dotted lines mark datasets that exceed 1k and 10k examples respectively.}\label{fig:data-loc-repair}
\end{figure}

Obtaining a dataset of multiple thousand real bug fixes can be challenging especially if the considered bug type occurs less frequent in open source projects. For this reason, we aim to explore how much the size of the fine-tuning dataset (the number of real bug fixes) influence the final performance of \realit. Therefore, we evaluate three variants of \realit~ pre-trained on 1x, 5x and 100x mutants per code snippet which we then fine-tune on several subsamples of our real bug fix datasets. We consider subsamples of 1\% (334), 3\% (996), 5\% (1.658), 10\% (3.314), 30\% (9.936), 50\% (16.559), 70\% (23.180), 90\% (29.802) of all bug fixes. To obtain more stable results, we fine-tune our models on three subsamples per sample size and evaluate fine-tuned models on our validation set. Averaged results for all three \realit~ variants fine-tuned on the generated subsamples are reported in Figure \ref{fig:data-loc-repair}.

\inlineheadingbf{Impact of the real bug fix dataset size} 
We can observe a clear trend that more real bug fixes lead to an improved performance across all \realit~ variants. This holds true even for small fine-tuning datasets of around 1000 (less than 5\% of the original dataset size) bug fixes. As reported by the authors of ManySStuBs4J~\cite{ManySStuBs4J} or PySStuBs~\cite{kamienski2021pysstubs}, real bug fix collections of this size can be obtained by mining the top 1000 most popular Github projects for the respective language.

\inlineheadingbf{Scaling real bug fixes vs. scaling mutants} 
Although we have seen that fine-tuning on more real bug fixes increases the performance, it is actually difficult to scale up the number of real bug fixes (as the number of publicly accessible projects to mine real bug fixes from is limited). In contrast, generating more mutants per code snippet is more cost effective. For example, to achieve the same performance gain obtained from scaling the number of mutants generated from 5x to 100x, we have to fine-tune on at least 10\% of our real bug fix dataset (3314 bugs). Still, although scaling the number of mutants is preferred, the scaling effect is however limited, as we have seen in RQ3. 

\section{Threats to validity}
Although a variety of learning-based bug localization and repair models have been developed in recent years, there does not
exist a unique setup for training and benchmarking these models which is universally accepted. Therefore, even though we implemented our baselines close to the reference implementation, the resulting trained models might behave differently than in the setup they were originally developed for. 
To still achieve comparable results, we designed our evaluation to replicate prior studies~\cite{hellendoorn2019global} on neural bug localization and repair models as close as possible. For example, we adopted the same publicly accessible Github corpus ETH Py150k, a similar architectural design and similar baselines as employed by Hellendoorn et al.~\cite{hellendoorn2019global}. To further support a wider range of bug types 
which allowed us to exploit the hand validated benchmark PyPIBugs, we adjusted the architecture and mutation process similar to Allamanis et al.~\cite{allamanis2021self}. 
Still, our evaluation results for the baseline algorithms (in Table \ref{tab:results}) are slightly different than the results of prior studies. 
For example, we found that while graph-based models such as GNNs and GREAT still perform better in localizing and repairing real bugs, Transformers show a surprisingly strong performance on mutants. Note however that Hellendoorn et al.~\cite{hellendoorn2019global} anticipated this result (even though they only evaluated on variable misuses) when the models are trained for a longer duration -- which we did by training on approximately 2.4x more training examples.
In contrast to the results of Allamanis et al.~\cite{allamanis2021self}, we observe that graph-based models underperform in our evaluation setup which we attribute to two main differences: (1) for a fair comparison, all models only have access to the function implementation without the implementation context which prohibits the computation of type related or call structure related information exploited by the graph-based models and (2) we trained all models on a different (potentially smaller) dataset. Although integrating the type of information and training on a larger dataset would potentially benefit all baselines, the performance ranking between architectures might differ. However, since our experiments showed that the performance gain due to training on real bug fixes is unique and the effect could not be replicated by training on mutants, we
expect that adapting our evaluation setup has little to no influence on our evaluation outcome.

\section{Related Work}
We discuss the most related previous or concurrent work that (1) tackle single token bug localization and repair with alternative training strategies, (2) exploit real bug fixes for automatic program repair or code mutations and (3) consider alternative pre-train-and-fine-tune techniques.

\inlineheadingbf{Single token bug localization and repair} 
The detection and repair of single token bugs have been explored in previous work~\cite{allamanis2017learning, pradel2018deepbugs, vasic2019neural, hellendoorn2019global, richter2022learning, patra2021semantic, allamanis2021self}. Allamanis et al.~\cite{allamanis2017learning} addressed the detection and repair of variable misuse bugs (which we also considered in this work) by representing programs as graphs. Vasic et al.~\cite{vasic2019neural} proposed a joint model for same task and Hellendoorn et al.~\cite{hellendoorn2019global} explored alternative program representations. These techniques all have in common that they do not learn from real bug fixes but from artificially mutated code. In contrast, while \realit~ employs a similar Transformer-based architecture as discussed by Hellendoorn et al.~\cite{hellendoorn2019global}, we showed that integrating real bug fixes in the training process is crucial for the localization and repair of real bugs.
More recent work~\cite{richter2022learning, patra2021semantic, allamanis2021self} also showed that the quality of training data is important for effective bug localization and repair. For example, employing a more realistic mutator~\cite{richter2022learning, patra2021semantic} (i.e. a mutator that is more likely to reproduce a real bug) or learning to inject hard to find bugs~\cite{allamanis2021self} can both improve the localization and repair performance. However, the integration of these approaches often increases the complexity by requiring to learn a mutation operator either prior and concurrent to the training process. With \realit, we showed that integrating real bug fixes, while relying on simpler and easier to implement mutation operators, can be sufficient to obtain a significant improvement in real bug localization and repair performance. 
Interestingly enough, a concurrent work~\cite{he2022distribution} also explored whether real bug fixes have an impact on the performance of learning-based bug detectors. Similar to \realit, their model is  pre-trained on mutants and then fine-tuned real bug fixes. Surprisingly, while the authors found that fine-tuning on real bug fixes improves precision (i.e. the number of correct programs classified as buggy), the recall (i.e. the number of real bugs detected and repaired) actually suffers. In contrast, we find that \realit~ improves the number of bugs detected and repaired significantly {\em while} training on real bug fixes can also decrease the false positive rate. We attribute the difference in our findings to significant differences to the \realit~ training process: (1) the number of real bug fixes we fine-tune on is several magnitudes larger, (2) the number of mutants generated per code snippet is significantly higher and (3) the distribution of buggy and bug-free programs is balanced both during pre-training and fine-tuning. We believe that especially (3) is key to success of \realit. Training on an unbalanced dataset (with significant more bug-free than buggy code) risks that the model defaults to not detecting a bug (which would result in a higher precision and lower recall by design). 

\inlineheadingbf{Learning from real bug fixes}
Real bug fixes are not only a valuable resource for learning to localize {\em and} repair single token bugs but they can also be effectively exploited for automatic program repair~\cite{tufano2019empirical, chen2019sequencer, li2020dlfix, lutellier2020coconut, bader2019getafix} or code mutations~\cite{tufano2019learning, patra2021semantic}. SequenceR~\cite{chen2019sequencer}, for example, learns from thousands of bug fixes to predict one-line bug patches. Dlfix~\cite{li2020dlfix} and CoCoNuT~\cite{lutellier2020coconut} improved the repair performance by proposing more effective learning strategies. In contrast to \realit, however, these techniques are designed to only repair a given program location and, hence, whether a program is buggy and where the bug has to be fixed has to be known beforehand. In addition, these techniques are often trained on real bug fixes only without considering mutants for the training process. We showed that learning from mutants is actually crucial to achieve high performing models. 
This observation is also supported by DrRepair~\cite{yasunaga2020graph} which showed that pre-training a repair models on artificial errors improved the repair performance on syntactic errors. Still, their approach rely on a compiler to detect this type of bugs. The type of single token bugs which we considered in this work are typically missed by a compiler.

Code mutation addresses the inverse problem of injecting a bug into a correct program. Tufano et al.~\cite{tufano2019learning} and Patra and Pradel~\cite{patra2021semantic} showed that bug fixes can be effectively leverage to learn code mutations by learning to replicate the original bug. Interestingly, Yasunaga et al.~\cite{yasunaga2021break} showed that repeatedly training a breaker and fixer that initially learn from real bug fixes but then provide training data for each other actually improves the performance of the fixer to repair syntactic bugs. While our work showed that real bug fixes are also crucial for bug detection, we believe that exploiting real bug fixes in the mutation process for training bug detection and repair models can be a promising direction for future work.

\inlineheadingbf{Pre-training and fine-tuning} Pre-training on large corpora of fuzzy data and then fine-tuning on a specific task with a smaller dataset has been shown to be highly successful in domains such as natural language processing~\cite{devlin-etal-2019-bert, raffel2020exploring}, image processing~\cite{kolesnikov2020big} and most recently programming language processing~\cite{feng2020codebert, kanade2019pre}. In contrast to \realit, these techniques are often pre-trained on a generic unrelated task where data is available before fine-tuning them on a specific task. \realit, however, is trained and fine-tuned with same architecture with largely the same objective of identifying and repairing buggy (or mutated) code. 

CuBERT~\cite{kanade2019pre} showed that pre-training on a generic corpus of Python code can improve the detection performance on variable misuses. However, the authors employed mutants instead of real bug fixes in the fine-tuning phase. In contrast, \realit~ is pre-trained on mutants and then fine-tuned on real bug fixes. A combination of these two approaches by applying \realit~ on top of a pre-trained model would be interesting and we leave this open for future work.

\section{Conclusion}
In this work, we explore the effect of training on real bug fixes and mutants on the performance of bug localization and repair models. For this, we propose \realit, a novel pre-train-and-fine-tune approach for learning to localize and repair bugs with Transformers. \realit~ can effectively utilize both mutants {\em and} real bug fixes during training by first pre-training on mutants and then fine-tuning on real bug fixes. Our evaluation on thousands of real bugs obtained from real Python projects showed that \realit~ can significantly improve the localization and repair of real bugs in contrast to models solely trained on mutants. In addition, our experiments showed (1) that pre-training on mutants plays an important role for achieving the performance level, (2) that mutants alone are however not sufficient to unlock the potential of \realit~ and (3) that a high number of real bug fixes is actually necessary for achieving a high performing model. 

Based on these observations, we see as future work the integration of more realistic data in the training process of neural bug localization and repair models. For example, training on more realistic mutants could boost the performance even before fine-tuning on real bug fixes. In addition, it might also be interesting to explore the effect of other -- even unrelated -- types of bug fixes on the training process of neural bug localization and repair approaches. Integrating more supported bug types also allows us to exploit more real bug fixes found in open source projects.

Finally, to conclude, \realit~ demonstrates that neural bug localization and repair models can effectively learn from developer mistakes, in form of real bug fixes, to localize and repair real bugs.


\bibliographystyle{ACM-Reference-Format}
\bibliography{references}

\clearpage

\appendix

\section{Model Architectures}
For our evaluation, we implemented our baselines in a common code base. All neural network modules
are implemented in PyTorch. In the following, we discuss the general architecture used for neural bug localization and repair and the design and hyperparameters individually for all baseline models.

\inlineheadingbf{General} All our models follow the general structure proposed by Hellendoorn et al.~\cite{hellendoorn2019global}. The architecture consists of an input module (for mapping tokens to vectors), a central encoding model and localization and repair head. For constructing our baselines, we change the central encoding model. The remaining structure remains the same (if not specified otherwise). For the input module, we use a BPE subtoken encoder with a vocabulary of 10k subtokens and embed each token by averaging its subtoken representation.

Similar to Allamanis et al.~\cite{allamanis2021self}, we employ dedicated heads for localization and repair. 

For localization, we use an architecture similar to pointer networks~\cite{merity2016pointer}. Given a program $\mathcal{T} = t_0, t_1, \dots, t_n$ and let $t_l$ be a potential bug location, we then compute the initial token embedding $e_l$ and the contextual vector representation $\mathbf{r}_l$ coming from the encoding model. Based on these representations, we compute a buggyness score for each potential bug location with a simple MLP:
\begin{equation*}
s_l = \mathbf{W}_2 \sigma(\mathbf{W}_1(\mathbf{r}_l || e_l || \mathbf{r}_l - e_l))
\end{equation*}
Here, $\mathbf{W}_2 \in \mathbb{R}^{1 \times d}$, $\mathbf{W}_1 \in \mathbb{R}^{d \times 3d}$ are learnable projections of the MLP. The intuition here is that the MLP should learn the correct token representation $\mathbf{r}_l$ which then would disagree with the initial token embedding $e_l$ if $t_l$ is buggy. We model the distribution $p_{loc}$ by a softmax over all buggyness scores.

Based on the same intuition used for localization, we designed our repair module. Given a potential bug location represented by $\mathbf{r}_l$, the repair module computes a repair score for all other tokens (represented by $\mathbf{r}_j$) similar to an attention mechanism:
\begin{equation*}
rep_{lj} = \frac{\mathbf{W}_q(\mathbf{r}_l) (\mathbf{W}_k (\mathbf{r}_j))^T}{\sqrt{d}}
\end{equation*}
Here, $\mathbf{W}_q \in \mathbb{R}^{d \times d}$, $\mathbf{W}_k \in \mathbb{R}^{d \times d}$ are learnable projections. To include an external vocabulary $V$, we represent each vocabulary entry a learnable vector $v_j \in \mathbb{R}^{d}$ and compute a repair score in a similar way:
\begin{equation*}
rep_{lj} = \frac{\mathbf{W}_q(\mathbf{r}_l) (v_j)^T}{\sqrt{d}}
\end{equation*}
Finally, $p_{repair}$ is computed by a softmax over all repair scores (token based and vocabulary based together).

We train all models using the Adam optimizer with learning rate $1e-4$ and a linear warm-up of 800 steps, additionally clipping gradient norms at 1.0 (0.5 for the GNN). Models are trained with weight decay of 0.1 for regularization. During training, we consider function implementations with up to 1024 tokens (1536 nodes for the GNN) and trained with minibatch sizes of up to 12.5K tokens (nodes).

\inlineheadingbf{RealiT} We follow the same architectural design for RealiT. As an encoding model, we employ a 6-layer Transformer encoder~\cite{devlin-etal-2019-bert}, a hidden size of 512, an intermediate size of 2048 and 8 attention heads. During training, we use a dropout regularization of 0.1. For encoding the positions of tokens, we employ relative position encoding \cite{shaw2018self} as we found that this performed better. A comparison of Transformer with and without relative position encoding can be found in Table \ref{tab:results} ("RealiT - without fine-tuning on real bug fixes" vs "Transformer").

\inlineheadingbf{GNN} For the graph neural network baseline, we followed the design of Allamanis et al.~\cite{allamanis2021self} as close as possible. We reimplemented the GNN based on the reference implementation provided by the authors. The GNN consists of 8 message propagation layers with a skip connection between the first and the fourth layer and between the fourth and the eights layer. The node hidden size is set to 256.

In addition, we also adapted the general architecture to match the reference implementation.
Instead of averaging the subtoken embeddings, we employ max pooling as the authors found that this performed better. In addition, we also reimplemented the same localization and repair head.

\inlineheadingbf{Remaining baselines} The remaining baselines employ the same hyperparameters as specified by Hellendoorn et al.~\cite{hellendoorn2019global}. The Transformer is a 6-layer encoder with absolute position embeddings (512 hidden size, 2048 intermediate size, 8 attention heads). GREAT uses a similar 6-layer architecture with the addition of an edge bias. The RNN is a 2-layer bidirectional recursive neural network with hidden size of 512.

\end{document}